\documentclass{JHEP3}
\usepackage{graphics}
\usepackage{amsfonts,amsbsy,latexsym,amssymb,amscd,amstext}
\usepackage{epsfig}
\usepackage{graphicx}
\title{Large $N$ reduction with the Twisted Eguchi-Kawai model}
\author{Antonio  Gonz\'alez-Arroyo $^{a}$ and Masanori Okawa $^{b}$ \\
  $^a$ Instituto de F\'{\i}sica Te\'orica UAM/CSIC,  C-8 \\
     and  Departamento de F\'{\i}sica Te\'orica, C-15 \\
                 Universidad Aut\'onoma de Madrid, \\
		 Cantoblanco E-28049--Madrid, Spain \\
		 $^b$ Graduate School of Science \\
		 Hiroshima University \\
		 Higashi-Hiroshima, Hiroshima 739-8526, Japan \\
		 
	    E-mail: \email{ antonio.gonzalez-arroyo@uam.es,
	okawa@sci.hiroshima-u.ac.jp}			       
				         }
						    
\abstract{
We examine the breaking of $Z_N$ symmetry recently reported for the
Twisted Eguchi-Kawai model (TEK).
We analyse the origin of this behaviour  and propose simple
modifications of twist and  lattice action  that could avoid the problem. Our
results show no sign of symmetry breaking and allow us to obtain
values of the large $N$ infinite volume string tension in agreement with
extrapolations from results based upon straightforward methods.}

	     
	                  \preprint{IFT-UAM/CSIC-10-32\\ FTUAM-2010-9
			  \\HUPD-1002}

\begin{document}
\maketitle

\section{Introduction}
The large $N$ limit of gauge theories continues to be a fascinating
source of simplifications of the complicated field theory dynamics,
which has yet defied our capacity for finding an analytic solution. A puzzling
old phenomenon is space-time reduction, namely the possibility put
forward by Eguchi and Kawai~\cite{EK} that the Physics of the large $N$
limit would be independent of the physical spatial volume. The
idea  originated  by comparing the loop
equations~\cite{makeenko-migdal}
of a lattice gauge theory on a $d$-dimensional torus of different sizes. 
The corresponding equations were independent of the lattice size provided 
the $Z_N^d$  symmetry of the infinite volume $d$-dimensional theory is preserved by the finite
volume theory. Taking the idea to the extreme, a one-point periodic lattice
Matrix model (the EK model) would encompass the dynamics of infinite
volume. Although, the idea seems to work at strong coupling, it was
observed that the symmetry and the reduction is lost for weak
coupling~\cite{BHN} in four dimensions. The same authors proposed a way
out, called Quenched Eguchi-Kawai model (QEK), in which the problem could be
avoided by freezing in the eigenvalues of the link matrices. The
present authors~\cite{AGAMO1}-\cite{AGAMO2} realized that the weak coupling behaviour of the
reduced model depends strongly on the boundary conditions while the
original Eguchi-Kawai proof does not. They proposed imposing twisted
boundary conditions in such a way as to guarantee that the symmetry
would be respected at weak coupling. The realization of this idea,
known as the twisted Eguchi-Kawai model (TEK), was studied analytically
and numerically and passed all tests with flying colours. Being more
effective in approaching  the large $N$ limit than the QEK model, it was
used both numerically and analytically for a better understanding of
large $N$ gauge theory~\cite{das}. 

It is interesting to recall how the TEK model is capable of recovering the
perturbative behaviour of the infinite volume theory. Indeed, by
expanding around the minimum action solutions, in a suitable SU(N) Lie
algebra basis, one obtains very similar  Feynman rules as for a
$L^4\equiv(\sqrt{N})^4$ lattice gauge theory. The momentum degrees of freedom
follow from the ones in the group. The propagators are exactly the
same as in the ordinary $L^4$ lattice theory, but the
vertices reflect the non-commutative nature of the group in that, in
addition to momentum conservation, there are momentum dependent phase
factors. These phases are basically the structure constants of the
group written in this basis. In Ref.~\cite{AGAMO2} the present authors
showed how phases cancel each other out  for planar diagrams, while they
remain for non-planar ones. Thus, all one needs to recover the
perturbative behaviour  of large $N$  gauge theory is that the phases
kill the contribution of the corresponding diagrams at large $N$. This 
was plausible since the phases oscillate more and more as we keep the
momenta fixed and take the large $N$ limit. Translating the mechanism
to the continuum~\cite{AGA-KA} led to non-commutative momenta and
Feynman rules which coincide with what later came to be known as
non-commutative quantum field theory~\cite{douglas}. Proof of
cancellation of the phases goes unchanged. 
The connection with non-commutative quantum field theory triggered 
a revival in the interest upon the TEK model, as a possible regulator
of the non-commutative gauge theory~\cite{ambjorn}.  

Recently, however, new results~\cite{TIMO}-\cite{TV}-\cite{Az} have shown that as $N$
is increased to higher values ($\ge 100$)  the TEK model shows a
pattern of symmetry breaking at intermediate couplings, with
pronounced hysteresis cycles. Although, the continuum limit  of the
theory sits at weak coupling, where symmetry is restored, the phase
transition point $\beta_c$ seems to move to higher and higher values of the
coupling  as $N$ grows. Since only loops of size smaller that
$L$ can be computed, the authors of Ref.~\cite{Az} claimed that 
their  physical size $La(\beta_c)$ shrinks to zero in the large $N$
limit, leaving no window for a continuum reduction. In parallel,
 evidence was given~\cite{sharpe} that the QEK model also
displays symmetry breaking at weak coupling.

 We refer that, on the positive side,  there have been important
 progress recently in
extending the reduction idea to gauge theories with fermions in the
adjoint representation~\cite{KUY}, where the fermions help in
stabilising the symmetry. Furthermore, a new approach to reduction has
been initiated by Narayanan, and Neuberger~\cite{NN} in which by
keeping both space and group degrees of freedom one can approach the 
large $N$ limit with optimal use of the degrees of freedom.

The purpose of this paper is to analyse the origin and nature of the
symmetry breaking on the TEK model  and  examine ways in which the
reduction idea can be restored for pure Yang-Mills gauge theory.

\section{Weak-coupling Analysis}
The TEK model is a model of $d$  SU(N) random matrices fluctuating with
a probability distribution derived from the
action  
\begin{equation}
S= b N  \sum_{\mu \nu} (N -z_{\mu \nu} {\rm Tr}(P_{\mu \nu})) 
\end{equation}
where the $z$ factors are elements of $\mathbf{Z}_N$ 
\begin{equation}
z_{\mu \nu}= \exp\{2 \pi i \frac{n_{\mu \nu}}{N}\} 
\end{equation}
and 
\begin{equation}
P_{\mu \nu} = U_\mu U_\nu U_\mu^\dagger U_\nu^\dagger 
\end{equation}
This model follows from considering SU(N) lattice gauge theory 
(with Wilson action) in a periodic lattice with twisted boundary 
conditions and collapsing the lattice to a single point. According 
to the proof of reduction\cite{EK}, this model should be equivalent
to the infinite volume lattice gauge theory at large $N$. This should be
so, irrespective on the value of $z_{\mu \nu}$. However, the proof of 
{\em reduction} relies on the hypothesis that the reduced model
respects the $Z_N^d$ symmetry of the large volume theory. The action is
obviously symmetric under  
\begin{equation}
\label{eq.symmetry} 
U_\mu \longrightarrow e^{ 2 \pi i p_\mu /N} U_\mu
\end{equation}
but this symmetry can be broken spontaneously, as we know this is
actually the case~\cite{BHN} for the original EK model ($z_{\mu \nu}=1$)
in 4 dimensions. From now on we will restrict ourselves for simplicity
to the four-dimensional case ($d=4$). 

There is a special interest in focusing upon the weak-coupling region
for two main reasons. The first, is that customarily the origin of
spontaneous symmetry breaking is the fact that the different classical
vacua (the absolute minima of the potential) are  not invariant under
the symmetry. This is actually the case for the EK model, and explains
why it fails to achieve reduction. The second reason, is that the
continuum limit of the theory lies at zero-coupling. Thus, the only
way in which we can dream of maintaining the reduction idea in the
continuum is by preserving the symmetry in the weak-coupling region. 

These considerations motivated us to examine the situation at weak
coupling and to realize that the choice of $z_{\mu \nu}$ has a
dramatic impact upon the classical vacua and symmetry
breaking~\cite{AGAMO1}-\cite{AGAMO2}. The
minimum action solution (called {\em twist-eaters}) $U_\mu=\Gamma_\mu$
can be made invariant under a combined transformation
Eq.~\ref{eq.symmetry} and a gauge rotation. It is impossible though to
preserve the full $Z_N^4$ of transformations. At most one can preserve 
a $Z_N^2$ subgroup. Guided by the requirement of treating all
directions equally, we proposed to take $N=L^2$ and the following 
{\em symmetric twist} choice of the antisymmetric twist tensor:
\begin{equation}
n_{\mu \nu}= k L \quad \quad {\rm for } \quad \mu < \nu 
\end{equation}
Our particular proposal was taking $k=1$, but indeed any integer which
is coprime with $L$ suffices~\footnote{values of $k$ other than one were
considered in the non-commutative geometry context and also in
Ref.~\cite{TIMO}.}. This twist tensor is by no means unique, 
but it is the most symmetric proposal, and the remaining symmetry group
can be written as $Z_L^4$ and allows the fulfilment of the reduction
concept as $L$ goes to infinity. Indeed, all numerical studies of the TEK 
model done at the time of our paper revealed that the symmetry was preserved 
at all values of the coupling. 

Since the symmetry cannot be broken to any order in perturbation
theory, the perturbative expansion of the TEK model must reproduce the one of
infinite volume gauge theory at infinite $L$. The way in which this is
achieved is fascinating, and anticipated by many years developments 
in non-commutative space-time field theories (for a review consult
Ref.~\cite{douglas}). Let us 
review here some aspects, generalising the formulas  of
Ref.~\cite{AGAMO2} to arbitrary $k$ coprime with $L$. 

The first step is to expand  the link variables around
twist-eaters: 
$ U_\mu= e^{i A_\mu} \Gamma_\mu $
and replace this expression into  $P_{\mu \nu}$:
\begin{equation}
z_{\mu \nu} P_{\mu \nu}= e^{i A_\mu} e^{i \delta_\mu A_\nu} e^{-i
\delta_\nu A_\mu} e^{-i
A_\nu} = e^{i G_{\mu \nu}}
\end{equation}
where
\begin{equation}
 \delta_\mu \phi \equiv  \Gamma_\mu \phi  \Gamma^{\dagger}_\mu \equiv D_\mu
\phi + \phi
\end{equation}
Obviously $G_{\mu \nu}$ is hermitian and given by the
Baker-Campbell-Hausdorff formula:
$$ G_{\mu \nu}= D_\mu A_\nu - D
_\nu A_\mu + O(A^2) $$
Our notation is chosen to stress  the resemblance with the ordinary 
field-strength tensor. And indeed the similarity goes beyond, since the 
operators $D_\mu$ turn out to commute among themselves and have the 
same spectrum as the ordinary lattice derivatives in an $L^4$ lattice:
\begin{equation}
D_\mu \lambda(q) = (e^{iq_\mu} -1) \lambda(q)
\end{equation}
with $q_\mu$ and integer multiple of $2 \pi/L$. The SU(N) matrices
$\lambda(q)$, whose explicit form we will not need, 
are the corresponding eigenvectors. Hence, perturbation theory propagators
will reproduce the standard  lattice propagator in an $L$ dimensional lattice. This gives us
information about the nature of the finite $L$ corrections to the
reduction picture. Of course, before reaching that step, one must handle 
zero-modes of the quadratic form in $A_\mu$. However, since the vacuum
is unique up to gauge transformations, one can get rid of them by an
appropriate gauge fixing. Gauge transformations adopt the familiar form
\begin{equation}
e^{i A_\mu} \longrightarrow \Omega e^{i A_\mu} (\delta_\mu \Omega)^\dagger
\end{equation}
if we conceive $\delta_\mu$ as the operator of displacement by one
lattice spacing in the $\mu$ direction. 

However, the reduced model at finite $L$ is  not simply equivalent 
to the ordinary lattice gauge model in a finite volume. The vertices
of the theory contain (in addition to momentum conservation delta
functions) momentum dependent phases, which (the antisymmetric part) are simply 
the values of the structure constants of the theory in the basis of eigenstates of
$D_\mu$: 
\begin{equation}
{\rm Tr}(\lambda(q) \lambda(p) \lambda(k)) = \delta(q+p+k)
e^{i\Phi(q,p)\theta_L}
\end{equation}
with $\Phi(q,p)=-\langle q,q \rangle -\langle p,p \rangle -\langle p,q 
\rangle$, up to additional terms depending on the normalization of the 
$\lambda(q)$ matrices. The symbol  $\langle p,q\rangle$ stands for the
quadratric form
$$ 
\langle p,q\rangle =p_0(q_1-q_2+q_3)+ p_1(q_2-q_3)+p_2q_3 $$
and 
\begin{equation}
\theta_L= \frac{L \bar{k}}{2 \pi} 
\end{equation}
where $\bar{k}$ is an integer, coprime with $L$, and defined through
the relation:
\begin{equation}
\label{defbark}
k \bar{k}= mL +1
\end{equation}
In fact, $\bar{k}$ and $\theta_L$ are the only places in which the
perturbative series depends on the choice of $k$. The momentum
dependent phases are crucial to recover the large $N$ dynamics since,
as proven in Ref~\cite{AGAMO2}-\cite{SQ}, 
the  phase factors from different vertices cancel out for  planar
diagrams.  For non-planar diagrams, the remaining phase factor will
oscillate more and more strongly as $L$ grows and so,  we argued,  
it will cancel it out. 

Some comments are in order. First, that we are implicitly assuming that
the large $N$ limit is taken on the lattice theory at fixed $b$, so
that one does not have to worry about ultraviolet divergences. Only
later, the continuum limit is taken by the customary scaling as
$b$ goes to infinity, with the large $N$ beta function. A different
story occurs if we want to take a {\em double scaling limit} in which we
keep the continuum non-commutativity parameter fixed $\theta=\theta_L
a^2(\beta)$, to obtain continuum non-commutative theories. Although,
very interesting by itself, we will not bother about it in the present
work, since it is not our main goal. 

Altogether, we have reviewed  our proof of reduction based on
perturbation theory, which serves to complement the loop equation
proof of Eguchi and Kawai. Focusing again upon  $k$ dependence, notice
that the suppression of non-planar diagrams  is bigger for large
values of $\theta_L$ and hence, of $\bar{k}$. 

This panorama remained stable for many years, until recent simulations done
at larger values of $N$ showed a pattern of $Z_L^4$ symmetry breaking at 
intermediate couplings~\cite{TIMO}-\cite{TV}-\cite{Az}. As a matter of fact, the authors 
of Ref.~\cite{TV} express doubts that the reduction applies for
physical sizes below $1/T_c$, and in  Ref.~\cite{Az} the conclusion
seems to be that the window of physical sizes where the symmetry 
is preserved shrinks to zero as $N$ grows. Motivated by these results
we decided to explore the situation in more detail. 

The authors of Ref.~\cite{TV} suggest that the origin of the symmetry
breaking could be due to other extrema of the TEK  action
functional~\cite{vanbaal}.  We follow the nomenclature used in those
references and refer to these extrema as {\em fluxons}. These extrema
are given by $U_\mu=\Gamma'_\mu$ such that the  plaquette values  are $Z_N$
elements:
\begin{equation}
\Gamma'_\mu \Gamma'_\nu= e^{-\frac{2 \pi i n'_{\mu \nu}}{N}}
\Gamma'_\nu \Gamma'_\mu
\end{equation}
Each fluxon has its characteristic pattern of symmetry breaking and a
set of products having possible non-vanishing traces. We call these products
{\em open paths} for obvious reasons. One particular example of fluxon
are the twist-eaters themselves ($n'_{\mu \nu}=n_{\mu \nu}$). For them
the open paths must have length proportional to $L$ in all directions. 
The extreme opposite case is that of singular {\em torons} (using the
terminology of Ref.~\cite{GAJKA}) having
$U_\mu=z_\mu \mathbf{I}$. In that case, the symmetry breaks down 
completely and all paths are open. 

As a matter of fact,  fluxons  are extrema of the TEK model for any
choice of twist  $n_{\mu \nu}$. This is a huge number of extrema. 
What the choice of twist determines is the action of each fluxon. 
However, as argued in Ref.~\cite{TV}, as $b$ decreases, 
entropy might well overcome the energy difference and the system can jump
to the vicinity of a  different fluxon, inducing symmetry breaking. Let
us examine the value of the action difference between a singular
toron and a twist eater:
\begin{equation}
\label{DeltaS}
\Delta S= 24 b N^2  \sin^2(\frac{\pi k}{L}) 
\end{equation}
As we take the large $N$ limit with $b$ and $k$ fixed, this difference
in action grows as  $N$. This, might not be enough if the entropy
difference is order $N^2$ (as is natural since that is the number of
degrees of freedom). 

The aforementioned problem seems to have a simple solution. One has to
take values of $k$ which are order $L$. Is this possible? A priori there 
seems to be no objection, since the perturbative proof of reduction
only requires $k$ to be coprime with $L$.  The only possible effect 
at finite $L$  was hidden in the value of $\theta_L
$. We will examine 
this point later. 

It is clear, from Eq.~\ref{DeltaS}, that the maximum action difference 
that we can obtain by changing $k$ is  $12bN^2$. Is this enough? If the entropy
difference is proportional to $N^2$ then there would be a weak
coupling region in which symmetry would be respected and a continuum
limit taken. The question then would be: 
What is then the minimum value of $b$ which can be explored?

Unfortunately, there is even a more worrisome situation if the entropy 
difference between the  twist-eater and the singular toron grows as 
$N^2 \log(N)$, as suggested by the analysis of Ref.~\cite{GAJKA}. 
The log behaviour comes from the vanishing of the linear term  in  
$G_{\mu \nu}$. Fluctuations are then quartic rather than quadratic.  

Nevertheless, if we study   fluctuations around the singular toron, we
obtain 
\begin{equation}
S= 24 b N^2  \sin^2(\frac{\pi k}{L}) + 6 b N  \cos(\frac{2 \pi k
}{L}) {\rm Tr}(G^2_{\mu \nu}) + \ldots
\end{equation}
Thus, we see that for $k/L >1/4$, the singular toron becomes unstable
(it becomes a local maximum). This fact applies as well to all other
torons. 

What about other fluxons? Could we fall into the same problem with other fluxons?
First of all, one must say that the singular toron is the most dangerous case,
because it has the largest number of quartic fluctuations. In no other
case can one have $N^2 \log N$ contributions. There are cases,
nonetheless, in which $N \log N$ contributions can appear. For that
reason  we have analysed all fluxons  to see  if we could get fluxons  having 
action differences of order one, and possessing short open paths. 
Indeed, this possibility cannot occur. Furthermore, if $k$ is order
$L$ the bounds get stronger and one can rule out the possibility that
an $N \log N$ entropy could win over the energy\footnote{To simplify
the proof  we assumed that $L$ is prime and $\bar{k}$ is not of order $L$.}.
The detailed analysis is lengthy and will be presented elsewhere\cite{AGAMO4}.

Hence, we conclude that a judicious choice of $k(L)$ could solve the
instabilities observed in the TEK model for $k=1$ and lead to a
reduced large $N$ limit. Our arguments are based on weak coupling
analysis and also explain why the $k=1$ failed. These arguments cannot
rule out differences of entropy of order $N^2$, but as mentioned earlier
this will leave a window of $b$ values where reduction will apply.
This can be important as a matter of principle even though the values 
of $b$ for large $N$ reduction to apply could be too large for all
practical purposes. For that reason,  it is important to realize that 
all the dangerous fluxons are not features of the continuum theory. 
Thus, their action can be changed appropriately by changing the lattice 
action of the TEK model, without affecting the continuum formulation in
the symmetry breaking phase. This  strategy is currently being studied by 
the present authors. 

Finally, we should comment, that for a faster approach to the large $N$  
infinite volume theory a large value of $\bar{k}$ is desirable. Indeed, the value of
$\bar{k}$ can also be chosen to grow proportionally to  $L$. To show
that all these conditions are not mutually exclusive, we give an example:
$k=\bar{k}=(4p -1)$ and  $L= (16p-8)$,  for any positive integer $p$.

\section{Numerical results}
Motivated by our previous arguments, we  studied the behaviour of the
TEK model for different values of $k$ by Monte Carlo methods. We
employed the Fabricius-Hahn heat-bath algorithm~\cite{FH}.
Our first goal was  to determine the value of $b_c(L,k)$ below which symmetry breaking is 
observed when starting  from cold initial configurations based on twist-eaters. 
Since large hysteresis have been reported, $b_c(L,k)$ would  actually
turn out to be  smaller than the phase transition point.  But for our purposes it is
admissible to do simulations in a metastable vacuum provided the
lifetimes are large enough for accumulating sufficient statistics.
The $k=1$ case has been studied previously by other
authors\cite{TIMO}-\cite{TV}-\cite{Az}, but we have extended the analysis to higher  
values of $L$. It is known that for $L\le 9$ there
is no evidence for symmetry breaking. Beyond, the traces of the link
variables develop  a non-zero value for $b<b_c(L,1)$. 
The  values of $b_c(L,1)$ are plotted as a function of $L$
in Fig.~\ref{fig1}. For presentation purposes we have omitted from the Figure
our largest group values: $b_c(23,1)=0.8075(25)$ and
$b_c(27,1)=1.1925(25)$.

The $k=2$ case was also addressed in Ref.~\cite{TIMO}, where it was
found that it also exhibits symmetry breaking for $L\ge 19$.  Here  we 
have extended this analysis  to much higher values of $L$.  The values of
$b_c(L,2)$ seem to match with those of $k=1$ if plotted as a function
of $L/k$ as seen in Fig.~1. Indeed, there is small downward shift of
0.0085, in addition to scaling, which we have  incorporated in the plot.
Scaling in $L/k$ (with the same shift) is also observed for 
$k=3$ although the range is limited, since there is no breaking for 
$L \le 28$. For $k=4$ we verified 
that symmetry remains unbroken up to (and including) $L=37$. For
$L$=39 and $b$ =0.34 we observed symmetry breaking, giving a
values of $b_c(L,4)$ which is  consistent with scaling. 
For $k=5,6,7$ we have
been unable to observe symmetry breaking. All our results point
towards a symmetry breaking curve $b_c(L,k)$ which scales
approximately with $L/k$ and is defined only for $L/k> 9-9.5$.
To show evidence of the symmetry restoration, we show in Fig.~\ref{fig2} 
the distribution of eigenvalues of the four  link variables for one of 
$L=23$, $k=7$, $b=0.37$ configurations, showing the characteristic structureless 
flat pattern.

%\begin{figure}
\FIGURE{
\label{fig1}
% GNUPLOT: LaTeX picture with Postscript
\begingroup
  \makeatletter
  \providecommand\color[2][]{%
    \GenericError{(gnuplot) \space\space\space\@spaces}{%
      Package color not loaded in conjunction with
      terminal option `colourtext'%
    }{See the gnuplot documentation for explanation.%
    }{Either use 'blacktext' in gnuplot or load the package
      color.sty in LaTeX.}%
    \renewcommand\color[2][]{}%
  }%
  \providecommand\includegraphics[2][]{%
    \GenericError{(gnuplot) \space\space\space\@spaces}{%
      Package graphicx or graphics not loaded%
    }{See the gnuplot documentation for explanation.%
    }{The gnuplot epslatex terminal needs graphicx.sty or graphics.sty.}%
    \renewcommand\includegraphics[2][]{}%
  }%
  \providecommand\rotatebox[2]{#2}%
  \@ifundefined{ifGPcolor}{%
    \newif\ifGPcolor
    \GPcolortrue
  }{}%
  \@ifundefined{ifGPblacktext}{%
    \newif\ifGPblacktext
    \GPblacktexttrue
  }{}%
  % define a \g@addto@macro without @ in the name:
  \let\gplgaddtomacro\g@addto@macro
  % define empty templates for all commands taking text:
  \gdef\gplbacktext{}%
  \gdef\gplfronttext{}%
  \makeatother
  \ifGPblacktext
    % no textcolor at all
    \def\colorrgb#1{}%
    \def\colorgray#1{}%
  \else
    % gray or color?
    \ifGPcolor
      \def\colorrgb#1{\color[rgb]{#1}}%
      \def\colorgray#1{\color[gray]{#1}}%
      \expandafter\def\csname LTw\endcsname{\color{white}}%
      \expandafter\def\csname LTb\endcsname{\color{black}}%
      \expandafter\def\csname LTa\endcsname{\color{black}}%
      \expandafter\def\csname LT0\endcsname{\color[rgb]{1,0,0}}%
      \expandafter\def\csname LT1\endcsname{\color[rgb]{0,1,0}}%
      \expandafter\def\csname LT2\endcsname{\color[rgb]{0,0,1}}%
      \expandafter\def\csname LT3\endcsname{\color[rgb]{1,0,1}}%
      \expandafter\def\csname LT4\endcsname{\color[rgb]{0,1,1}}%
      \expandafter\def\csname LT5\endcsname{\color[rgb]{1,1,0}}%
      \expandafter\def\csname LT6\endcsname{\color[rgb]{0,0,0}}%
      \expandafter\def\csname LT7\endcsname{\color[rgb]{1,0.3,0}}%
      \expandafter\def\csname LT8\endcsname{\color[rgb]{0.5,0.5,0.5}}%
    \else
      % gray
      \def\colorrgb#1{\color{black}}%
      \def\colorgray#1{\color[gray]{#1}}%
      \expandafter\def\csname LTw\endcsname{\color{white}}%
      \expandafter\def\csname LTb\endcsname{\color{black}}%
      \expandafter\def\csname LTa\endcsname{\color{black}}%
      \expandafter\def\csname LT0\endcsname{\color{black}}%
      \expandafter\def\csname LT1\endcsname{\color{black}}%
      \expandafter\def\csname LT2\endcsname{\color{black}}%
      \expandafter\def\csname LT3\endcsname{\color{black}}%
      \expandafter\def\csname LT4\endcsname{\color{black}}%
      \expandafter\def\csname LT5\endcsname{\color{black}}%
      \expandafter\def\csname LT6\endcsname{\color{black}}%
      \expandafter\def\csname LT7\endcsname{\color{black}}%
      \expandafter\def\csname LT8\endcsname{\color{black}}%
    \fi
  \fi
  \setlength{\unitlength}{0.0500bp}%
  \begin{picture}(7200.00,5040.00)%
    \gplgaddtomacro\gplbacktext{%
      \csname LTb\endcsname%
      \put(1342,704){\makebox(0,0)[r]{\strut{} 0.3}}%
      \put(1342,1383){\makebox(0,0)[r]{\strut{} 0.35}}%
      \put(1342,2061){\makebox(0,0)[r]{\strut{} 0.4}}%
      \put(1342,2740){\makebox(0,0)[r]{\strut{} 0.45}}%
      \put(1342,3419){\makebox(0,0)[r]{\strut{} 0.5}}%
      \put(1342,4097){\makebox(0,0)[r]{\strut{} 0.55}}%
      \put(1342,4776){\makebox(0,0)[r]{\strut{} 0.6}}%
      \put(1965,484){\makebox(0,0){\strut{} 10}}%
      \put(2946,484){\makebox(0,0){\strut{} 12}}%
      \put(3927,484){\makebox(0,0){\strut{} 14}}%
      \put(4908,484){\makebox(0,0){\strut{} 16}}%
      \put(5889,484){\makebox(0,0){\strut{} 18}}%
      \put(6870,484){\makebox(0,0){\strut{} 20}}%
      \put(440,2740){\rotatebox{90}{\makebox(0,0){\strut{}$b_c(L,k)$}}}%
      \put(4172,154){\makebox(0,0){\strut{}$L/k$}}%
    }%
    \gplgaddtomacro\gplfronttext{%
      \csname LTb\endcsname%
      \put(2266,4603){\makebox(0,0)[r]{\strut{}$k=1$}}%
      \csname LTb\endcsname%
      \put(2266,4383){\makebox(0,0)[r]{\strut{}$k=2$}}%
      \csname LTb\endcsname%
      \put(2266,4163){\makebox(0,0)[r]{\strut{}$k=3$}}%
    }%
    \gplbacktext
    \put(0,0){\includegraphics{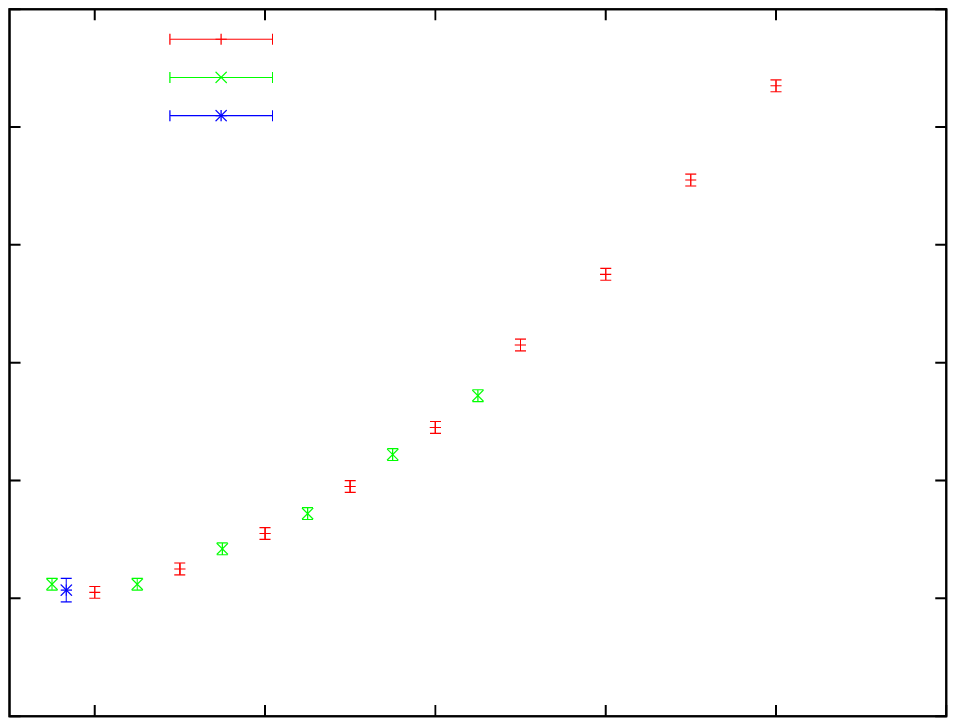}}%
    \gplfronttext
  \end{picture}%
\endgroup

\caption{The values of $b_c(L,k)$ are plotted as a function of $L/k$.
The values for $k=2$ and $k=3$ are shifted upwards by 0.0085 to
highlight scaling. The values of $k=1$ are from Ref.\cite{Az}.}
}
%\end{figure}

Scaling is not surprising if the dominant breaking mechanism is 
through transitions to the singular torons, since their action divided
by $N^2$, is indeed a function of $L/k$. Actually, we found that a
nice order parameter is the expectation value of the $\mu-\nu$
plaquettes formed by smeared link variables. Fluctuations are reduced 
considerably by smearing, and the mean value is centred at zero in the unbroken
phases, while for $k=1$ and $b<b(L,1)$ moves to the position
corresponding to torons. This is indeed, a confirmation that torons
are dominating the path integral in that region. 
\FIGURE{
%\begin{figure}
\label{fig2}
% GNUPLOT: LaTeX picture with Postscript
\begingroup
  \makeatletter
  \providecommand\color[2][]{%
    \GenericError{(gnuplot) \space\space\space\@spaces}{%
      Package color not loaded in conjunction with
      terminal option `colourtext'%
    }{See the gnuplot documentation for explanation.%
    }{Either use 'blacktext' in gnuplot or load the package
      color.sty in LaTeX.}%
    \renewcommand\color[2][]{}%
  }%
  \providecommand\includegraphics[2][]{%
    \GenericError{(gnuplot) \space\space\space\@spaces}{%
      Package graphicx or graphics not loaded%
    }{See the gnuplot documentation for explanation.%
    }{The gnuplot epslatex terminal needs graphicx.sty or graphics.sty.}%
    \renewcommand\includegraphics[2][]{}%
  }%
  \providecommand\rotatebox[2]{#2}%
  \@ifundefined{ifGPcolor}{%
    \newif\ifGPcolor
    \GPcolortrue
  }{}%
  \@ifundefined{ifGPblacktext}{%
    \newif\ifGPblacktext
    \GPblacktexttrue
  }{}%
  % define a \g@addto@macro without @ in the name:
  \let\gplgaddtomacro\g@addto@macro
  % define empty templates for all commands taking text:
  \gdef\gplbacktext{}%
  \gdef\gplfronttext{}%
  \makeatother
  \ifGPblacktext
    % no textcolor at all
    \def\colorrgb#1{}%
    \def\colorgray#1{}%
  \else
    % gray or color?
    \ifGPcolor
      \def\colorrgb#1{\color[rgb]{#1}}%
      \def\colorgray#1{\color[gray]{#1}}%
      \expandafter\def\csname LTw\endcsname{\color{white}}%
      \expandafter\def\csname LTb\endcsname{\color{black}}%
      \expandafter\def\csname LTa\endcsname{\color{black}}%
      \expandafter\def\csname LT0\endcsname{\color[rgb]{1,0,0}}%
      \expandafter\def\csname LT1\endcsname{\color[rgb]{0,1,0}}%
      \expandafter\def\csname LT2\endcsname{\color[rgb]{0,0,1}}%
      \expandafter\def\csname LT3\endcsname{\color[rgb]{1,0,1}}%
      \expandafter\def\csname LT4\endcsname{\color[rgb]{0,1,1}}%
      \expandafter\def\csname LT5\endcsname{\color[rgb]{1,1,0}}%
      \expandafter\def\csname LT6\endcsname{\color[rgb]{0,0,0}}%
      \expandafter\def\csname LT7\endcsname{\color[rgb]{1,0.3,0}}%
      \expandafter\def\csname LT8\endcsname{\color[rgb]{0.5,0.5,0.5}}%
    \else
      % gray
      \def\colorrgb#1{\color{black}}%
      \def\colorgray#1{\color[gray]{#1}}%
      \expandafter\def\csname LTw\endcsname{\color{white}}%
      \expandafter\def\csname LTb\endcsname{\color{black}}%
      \expandafter\def\csname LTa\endcsname{\color{black}}%
      \expandafter\def\csname LT0\endcsname{\color{black}}%
      \expandafter\def\csname LT1\endcsname{\color{black}}%
      \expandafter\def\csname LT2\endcsname{\color{black}}%
      \expandafter\def\csname LT3\endcsname{\color{black}}%
      \expandafter\def\csname LT4\endcsname{\color{black}}%
      \expandafter\def\csname LT5\endcsname{\color{black}}%
      \expandafter\def\csname LT6\endcsname{\color{black}}%
      \expandafter\def\csname LT7\endcsname{\color{black}}%
      \expandafter\def\csname LT8\endcsname{\color{black}}%
    \fi
  \fi
  \setlength{\unitlength}{0.0500bp}%
  \begin{picture}(7200.00,5040.00)%
    \gplgaddtomacro\gplbacktext{%
      \csname LTb\endcsname%
      \put(1078,704){\makebox(0,0)[r]{\strut{} 0}}%
      \put(1078,1383){\makebox(0,0)[r]{\strut{} 5}}%
      \put(1078,2061){\makebox(0,0)[r]{\strut{} 10}}%
      \put(1078,2740){\makebox(0,0)[r]{\strut{} 15}}%
      \put(1078,3419){\makebox(0,0)[r]{\strut{} 20}}%
      \put(1078,4097){\makebox(0,0)[r]{\strut{} 25}}%
      \put(1078,4776){\makebox(0,0)[r]{\strut{} 30}}%
      \put(1387,484){\makebox(0,0){\strut{}-3}}%
      \put(2271,484){\makebox(0,0){\strut{}-2}}%
      \put(3156,484){\makebox(0,0){\strut{}-1}}%
      \put(4040,484){\makebox(0,0){\strut{} 0}}%
      \put(4924,484){\makebox(0,0){\strut{} 1}}%
      \put(5809,484){\makebox(0,0){\strut{} 2}}%
      \put(6693,484){\makebox(0,0){\strut{} 3}}%
      \put(440,2740){\rotatebox{90}{\makebox(0,0){\strut{}$N_{\rm \small eigen}$}}}%
      \put(4040,154){\makebox(0,0){\strut{}$\theta$}}%
    }%
    \gplgaddtomacro\gplfronttext{%
      \csname LTb\endcsname%
      \put(1474,1537){\makebox(0,0)[r]{\strut{}0}}%
      \csname LTb\endcsname%
      \put(1474,1317){\makebox(0,0)[r]{\strut{}1}}%
      \csname LTb\endcsname%
      \put(1474,1097){\makebox(0,0)[r]{\strut{}2}}%
      \csname LTb\endcsname%
      \put(1474,877){\makebox(0,0)[r]{\strut{}3}}%
    }%
    \gplbacktext
    \put(0,0){\includegraphics{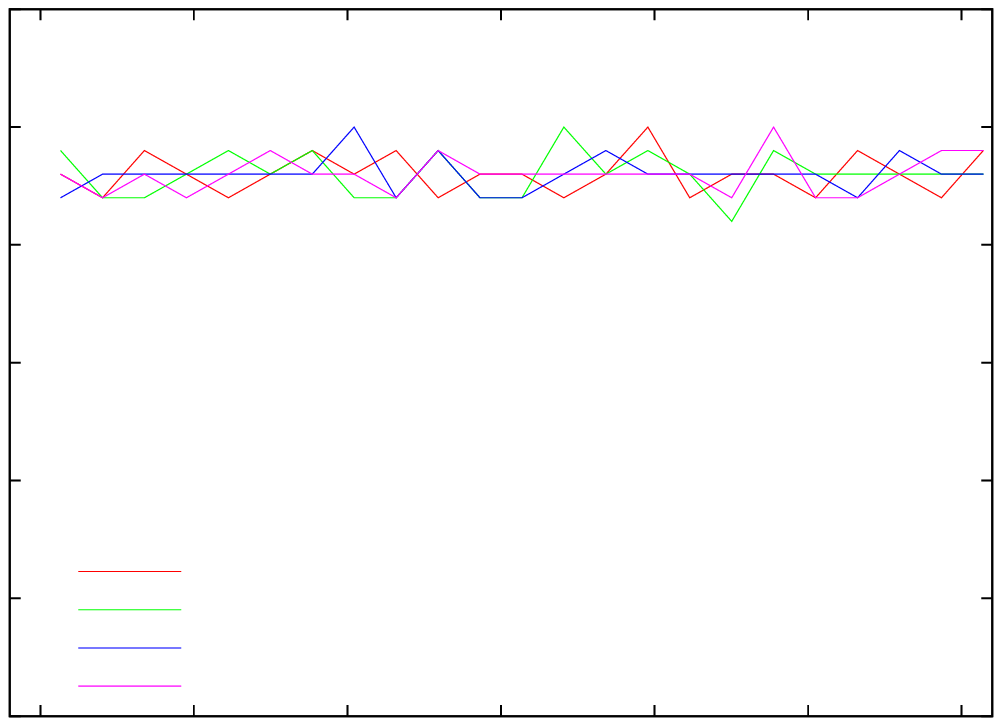}}%
    \gplfronttext
  \end{picture}%
\endgroup

\caption{We display a histogram showing the distribution of angles
from the eigenvalues $e^{i \theta}$ of  the
link matrices for a typical $L=23$ $k=7$ configuration at $b=0.37$.}
%\end{figure}
}
Scaling of $b_c(L,k)$ with $L/k$ might lead to worries related to the
possibility that  traces of open paths of approximate length $L/k$ acquire a
non-zero value. If that was the case, the advantages of exploring
higher values of $b$ with higher values of $k$ would be lost.
Indeed, the fluctuations of traces of open paths can be computed
in perturbation theory and depend on the choice of $k$. To leading
order one has 
\begin{equation}
\frac{1}{N^2}\langle |{\rm Tr}(U_0^{m_0}\cdots U_3^{m_3})|^2 \rangle
\propto \left(\sum_{\mu}\sin^2\left(\frac{ \pi  \sum_\rho n_{\mu \rho}
m_\rho}{N}\right)\right)
\end{equation}
Notice that paths having lengths of order $L/k$ have the largest
fluctuations. The risk is that these fluctuations could end up
destabilising the system and leading to symmetry breaking. To rule out
this possibility we analysed with particular detail all traces having
segments of length approximately equal to $L/k$. In all cases the
values of all traces were seen to oscillate with larger or smaller
amplitude around zero. 

Thus, our numerical results confirm the reasoning based on our weak
coupling analysis in what respects toron dominance, and  allows us 
to define a window of good choices of $k$. Our results point to  $k>L/9$ as
the region of $k$ values necessary to avoid symmetry breaking, as was 
first conjectured in Ref.~\cite{TIMO}.

At high $k$ values, signs of symmetry breaking have been observed in
the series of $k$ values $k=(L-1)/2$ with odd $L \ge 17$.  In fact,
at $L=17$ and $k=8$, Ishikawa and Okawa found large values of
$Tr(U^2_\mu)$ in their unpublished work.  We also found the same phenomena
at $L=19$, $21$ and $23$.  We, then,  made a very long run at $L=17$
and $k=8$ of 20000 sweeps.  Our results exhibit large autocorrelation
times in which the values of $Tr(U^2_\mu)$ are changing, so we cannot
conclude that we have symmetry breaking.  Furthermore, the smeared
plaquettes exclude that this is due to toron dominance.  It is
interesting to notice that the common feature of these cases is that
$|{\bar k} \bmod L|=2$ and considered previously in connection with the
non-commutative field theories [9].

These observations, together with the  lack of a formal proof that
the symmetry  will be respected at all values of the coupling and 
larger values of $L$, suggested us to adopt a more practical attitude,
and ask ourselves if the TEK model can, at the present stage, provide useful
results about the large $N$ limit of Yang-Mills theory. The rest of
this section will be devoted to showing that this is indeed the case.

Thus, we set ourselves to the task of exploring the model at the
largest  values of $L$ at which our present computer resources
allow us to collect  enough statistics.   We chose two values
$L=17$ and $L=23$. This  will  enable  us to quantify the $L$ dependence 
of our results.  In the first case, we chose $k=5$ and in the second
case $k=7$, satisfying our preferred choice criteria. 
For the $L=23$ case we generated 225 Monte Carlo configurations
separated by 20 sweeps, after discarding the initial 1500 sweeps,
for $b=0.36$ and $b=0.37$.  A sweep is defined by one heat-bath update
followed by five overrelaxation updates. Having two values of $b$ is also useful,
since it will allow us to analyse  scaling of our results. At $L=17$
we  generated 1025 configurations at $b=0.36$ with the same separation and
discarded initial sweeps.     First of all, notice that with our choice of $b$, 
we are exploring  physical sizes much beyond $1/T_c$,  avoiding the worries of
Ref.~\cite{TV}. One can use the estimates of $a(b=0.36) T_c$ extracted from
Ref.~\cite{LTW},  to compute the value of $l_{\rm \tiny phys} T_c$. For $L=23$ 
and $b=0.36$  this quantity is 2.75, well inside the confinement region.

Finally, we have attempted to extract physical results on the string
tension from our data. This determination is an update of our pioneer study of
string tension from reduced models~\cite{AGAMO3}-\cite{das}.  Recently, other
determinations based on the continuum reduction idea have
appeared~\cite{KN}. For that purpose, 
in each run we computed the expectation values of Wilson loops of all
sizes up to $\frac{(L-1)}{2}
\times \frac{(L-1)}{2}$. The signal obtained for large loops is  small
and the results very noisy. To obtain a better signal to background
ratio, we applied  $n_s$ smearing steps  to our link variables and recomputed
the loops with them. The smearing algorithm is 
\begin{equation}
U_\mu'= {\cal U} \left[ U_\mu + c \sum_{\nu \ne \mu} (z_{\nu \mu} U_\nu U_\mu
U_\nu^\dagger + z_{\mu \nu} U_\nu^\dagger U_\mu U_\nu)\right]   
\end{equation}
where ${\cal U}$ stands for the operator that projects onto unitary
matrices. Most of our results were obtained for $c=0.1$. For the
$L=17$ simulation  we also used  $c=0.3$ and found that  both values 
scale approximately  with the variable $fn_s$
of Ref.~\cite{NN2} ($f=6c/(1+6c)$). Noise has dropped considerably 
at $fn_s=4$. 

In order to determine the string tension we computed Creutz ratios  
\begin{equation}
\kappa(R,T)=
-\log\left(\frac{W(R,T)W(R-1,T-1)}{W(R,T-1)W(R-1,T)}\right) 
\end{equation}
and we fitted those having $R,T\ge 4$ to a formula of the type:
\begin{equation}
\label{kappa}
\kappa(R,T)= \sigma - \gamma
\left(\frac{1}{R(R-1)}+\frac{1}{T(T-1)}\right) 
\end{equation}
This dependence is obtained by plugging  the leading corrections expected in
perturbation theory for Wilson loops into the Creutz ratios formula. 
This method was used earlier, in a different context, with very good
results~\cite{GAM}. Actually, it also works very well in our case.
To show the quality of the fit, we display in Fig.~\ref{fig3} 
the value of $\sigma(R,T)$. The plotted
quantity is defined as 
\begin{equation}
\label{sigma}
\sigma(R,T)= \kappa(R,T) + \gamma \left(\frac{1}{R(R-1)}+\frac{1}{T(T-1)}\right)
\end{equation}
obtained from different Creutz ratios after
subtracting the correction proportional to $\gamma$ obtained from the
fit, as a function of the area $RT$. For a good fit, it should be flat, and its value
provides an estimate of the string tension.  This study has been  done for
all number of smearing steps with consistent results. As mentioned
previously best fits are obtained for $fn_s$ in the range $4-8$. 
An example is given in Fig.~\ref{fig3} for the results at $b=0.37$ and
$L=23$ after 15 smearing steps.  It is peculiar that the values
obtained for $\gamma$ were of the same sign and similar  magnitude to
those predicted by the universal string theory behaviour~\cite{LSW}. 
To appreciate the dependence of our results upon smearing, we 
display in Fig.~\ref{fig4} the list of string tensions obtained as a
function of the number of smearing steps for the $L=23$, $b=0.36$ case.   

\FIGURE{
%\begin{figure}
\label{fig3}
% GNUPLOT: LaTeX picture with Postscript
\begingroup
  \makeatletter
  \providecommand\color[2][]{%
    \GenericError{(gnuplot) \space\space\space\@spaces}{%
      Package color not loaded in conjunction with
      terminal option `colourtext'%
    }{See the gnuplot documentation for explanation.%
    }{Either use 'blacktext' in gnuplot or load the package
      color.sty in LaTeX.}%
    \renewcommand\color[2][]{}%
  }%
  \providecommand\includegraphics[2][]{%
    \GenericError{(gnuplot) \space\space\space\@spaces}{%
      Package graphicx or graphics not loaded%
    }{See the gnuplot documentation for explanation.%
    }{The gnuplot epslatex terminal needs graphicx.sty or graphics.sty.}%
    \renewcommand\includegraphics[2][]{}%
  }%
  \providecommand\rotatebox[2]{#2}%
  \@ifundefined{ifGPcolor}{%
    \newif\ifGPcolor
    \GPcolortrue
  }{}%
  \@ifundefined{ifGPblacktext}{%
    \newif\ifGPblacktext
    \GPblacktexttrue
  }{}%
  % define a \g@addto@macro without @ in the name:
  \let\gplgaddtomacro\g@addto@macro
  % define empty templates for all commands taking text:
  \gdef\gplbacktext{}%
  \gdef\gplfronttext{}%
  \makeatother
  \ifGPblacktext
    % no textcolor at all
    \def\colorrgb#1{}%
    \def\colorgray#1{}%
  \else
    % gray or color?
    \ifGPcolor
      \def\colorrgb#1{\color[rgb]{#1}}%
      \def\colorgray#1{\color[gray]{#1}}%
      \expandafter\def\csname LTw\endcsname{\color{white}}%
      \expandafter\def\csname LTb\endcsname{\color{black}}%
      \expandafter\def\csname LTa\endcsname{\color{black}}%
      \expandafter\def\csname LT0\endcsname{\color[rgb]{1,0,0}}%
      \expandafter\def\csname LT1\endcsname{\color[rgb]{0,1,0}}%
      \expandafter\def\csname LT2\endcsname{\color[rgb]{0,0,1}}%
      \expandafter\def\csname LT3\endcsname{\color[rgb]{1,0,1}}%
      \expandafter\def\csname LT4\endcsname{\color[rgb]{0,1,1}}%
      \expandafter\def\csname LT5\endcsname{\color[rgb]{1,1,0}}%
      \expandafter\def\csname LT6\endcsname{\color[rgb]{0,0,0}}%
      \expandafter\def\csname LT7\endcsname{\color[rgb]{1,0.3,0}}%
      \expandafter\def\csname LT8\endcsname{\color[rgb]{0.5,0.5,0.5}}%
    \else
      % gray
      \def\colorrgb#1{\color{black}}%
      \def\colorgray#1{\color[gray]{#1}}%
      \expandafter\def\csname LTw\endcsname{\color{white}}%
      \expandafter\def\csname LTb\endcsname{\color{black}}%
      \expandafter\def\csname LTa\endcsname{\color{black}}%
      \expandafter\def\csname LT0\endcsname{\color{black}}%
      \expandafter\def\csname LT1\endcsname{\color{black}}%
      \expandafter\def\csname LT2\endcsname{\color{black}}%
      \expandafter\def\csname LT3\endcsname{\color{black}}%
      \expandafter\def\csname LT4\endcsname{\color{black}}%
      \expandafter\def\csname LT5\endcsname{\color{black}}%
      \expandafter\def\csname LT6\endcsname{\color{black}}%
      \expandafter\def\csname LT7\endcsname{\color{black}}%
      \expandafter\def\csname LT8\endcsname{\color{black}}%
    \fi
  \fi
  \setlength{\unitlength}{0.0500bp}%
  \begin{picture}(7200.00,5040.00)%
    \gplgaddtomacro\gplbacktext{%
      \csname LTb\endcsname%
      \put(1342,704){\makebox(0,0)[r]{\strut{} 0}}%
      \put(1342,1518){\makebox(0,0)[r]{\strut{} 0.02}}%
      \put(1342,2333){\makebox(0,0)[r]{\strut{} 0.04}}%
      \put(1342,3147){\makebox(0,0)[r]{\strut{} 0.06}}%
      \put(1342,3962){\makebox(0,0)[r]{\strut{} 0.08}}%
      \put(1342,4776){\makebox(0,0)[r]{\strut{} 0.1}}%
      \put(1474,484){\makebox(0,0){\strut{} 0}}%
      \put(2245,484){\makebox(0,0){\strut{} 10}}%
      \put(3016,484){\makebox(0,0){\strut{} 20}}%
      \put(3787,484){\makebox(0,0){\strut{} 30}}%
      \put(4557,484){\makebox(0,0){\strut{} 40}}%
      \put(5328,484){\makebox(0,0){\strut{} 50}}%
      \put(6099,484){\makebox(0,0){\strut{} 60}}%
      \put(6870,484){\makebox(0,0){\strut{} 70}}%
      \put(440,2740){\rotatebox{90}{\makebox(0,0){\strut{}$\sigma(R,T)$}}}%
      \put(4172,154){\makebox(0,0){\strut{}AREA}}%
    }%
    \gplgaddtomacro\gplfronttext{%
    }%
    \gplbacktext
    \put(0,0){\includegraphics{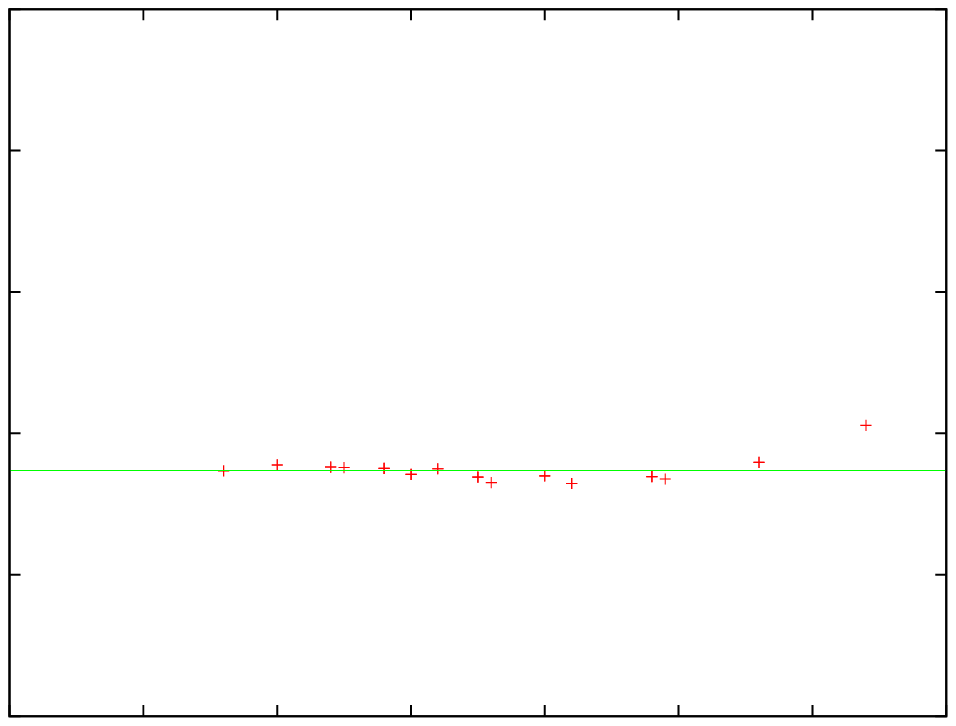}}%
    \gplfronttext
  \end{picture}%
\endgroup

\caption{We show the quantity $\sigma(R,T)$ defined in Eq.~\ref{sigma}
as a function of the area=$R\times T$. The horizontal line corresponds
to the string tension obtained from the fit. Data are for  the
$L=23$, $k=7$, $b=0.37$ configuration after 15 smearing steps.}
%\end{figure}
}

\FIGURE{
%\begin{figure}
\label{fig4}
% GNUPLOT: LaTeX picture with Postscript
\begingroup
  \makeatletter
  \providecommand\color[2][]{%
    \GenericError{(gnuplot) \space\space\space\@spaces}{%
      Package color not loaded in conjunction with
      terminal option `colourtext'%
    }{See the gnuplot documentation for explanation.%
    }{Either use 'blacktext' in gnuplot or load the package
      color.sty in LaTeX.}%
    \renewcommand\color[2][]{}%
  }%
  \providecommand\includegraphics[2][]{%
    \GenericError{(gnuplot) \space\space\space\@spaces}{%
      Package graphicx or graphics not loaded%
    }{See the gnuplot documentation for explanation.%
    }{The gnuplot epslatex terminal needs graphicx.sty or graphics.sty.}%
    \renewcommand\includegraphics[2][]{}%
  }%
  \providecommand\rotatebox[2]{#2}%
  \@ifundefined{ifGPcolor}{%
    \newif\ifGPcolor
    \GPcolortrue
  }{}%
  \@ifundefined{ifGPblacktext}{%
    \newif\ifGPblacktext
    \GPblacktexttrue
  }{}%
  % define a \g@addto@macro without @ in the name:
  \let\gplgaddtomacro\g@addto@macro
  % define empty templates for all commands taking text:
  \gdef\gplbacktext{}%
  \gdef\gplfronttext{}%
  \makeatother
  \ifGPblacktext
    % no textcolor at all
    \def\colorrgb#1{}%
    \def\colorgray#1{}%
  \else
    % gray or color?
    \ifGPcolor
      \def\colorrgb#1{\color[rgb]{#1}}%
      \def\colorgray#1{\color[gray]{#1}}%
      \expandafter\def\csname LTw\endcsname{\color{white}}%
      \expandafter\def\csname LTb\endcsname{\color{black}}%
      \expandafter\def\csname LTa\endcsname{\color{black}}%
      \expandafter\def\csname LT0\endcsname{\color[rgb]{1,0,0}}%
      \expandafter\def\csname LT1\endcsname{\color[rgb]{0,1,0}}%
      \expandafter\def\csname LT2\endcsname{\color[rgb]{0,0,1}}%
      \expandafter\def\csname LT3\endcsname{\color[rgb]{1,0,1}}%
      \expandafter\def\csname LT4\endcsname{\color[rgb]{0,1,1}}%
      \expandafter\def\csname LT5\endcsname{\color[rgb]{1,1,0}}%
      \expandafter\def\csname LT6\endcsname{\color[rgb]{0,0,0}}%
      \expandafter\def\csname LT7\endcsname{\color[rgb]{1,0.3,0}}%
      \expandafter\def\csname LT8\endcsname{\color[rgb]{0.5,0.5,0.5}}%
    \else
      % gray
      \def\colorrgb#1{\color{black}}%
      \def\colorgray#1{\color[gray]{#1}}%
      \expandafter\def\csname LTw\endcsname{\color{white}}%
      \expandafter\def\csname LTb\endcsname{\color{black}}%
      \expandafter\def\csname LTa\endcsname{\color{black}}%
      \expandafter\def\csname LT0\endcsname{\color{black}}%
      \expandafter\def\csname LT1\endcsname{\color{black}}%
      \expandafter\def\csname LT2\endcsname{\color{black}}%
      \expandafter\def\csname LT3\endcsname{\color{black}}%
      \expandafter\def\csname LT4\endcsname{\color{black}}%
      \expandafter\def\csname LT5\endcsname{\color{black}}%
      \expandafter\def\csname LT6\endcsname{\color{black}}%
      \expandafter\def\csname LT7\endcsname{\color{black}}%
      \expandafter\def\csname LT8\endcsname{\color{black}}%
    \fi
  \fi
  \setlength{\unitlength}{0.0500bp}%
  \begin{picture}(7200.00,5040.00)%
    \gplgaddtomacro\gplbacktext{%
      \csname LTb\endcsname%
      \put(1342,704){\makebox(0,0)[r]{\strut{} 0}}%
      \put(1342,1518){\makebox(0,0)[r]{\strut{} 0.02}}%
      \put(1342,2333){\makebox(0,0)[r]{\strut{} 0.04}}%
      \put(1342,3147){\makebox(0,0)[r]{\strut{} 0.06}}%
      \put(1342,3962){\makebox(0,0)[r]{\strut{} 0.08}}%
      \put(1342,4776){\makebox(0,0)[r]{\strut{} 0.1}}%
      \put(2553,484){\makebox(0,0){\strut{} 5}}%
      \put(3902,484){\makebox(0,0){\strut{} 10}}%
      \put(5251,484){\makebox(0,0){\strut{} 15}}%
      \put(6600,484){\makebox(0,0){\strut{} 20}}%
      \put(440,2740){\rotatebox{90}{\makebox(0,0){\strut{}$\sigma$}}}%
      \put(4172,154){\makebox(0,0){\strut{}Number of smearing steps}}%
    }%
    \gplgaddtomacro\gplfronttext{%
    }%
    \gplbacktext
    \put(0,0){\includegraphics{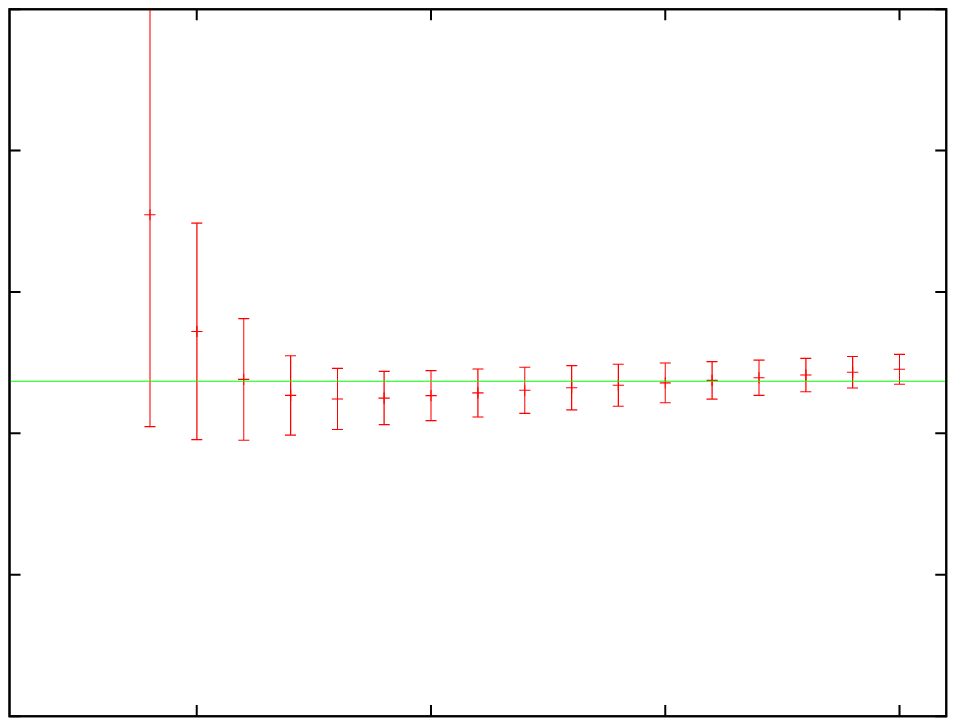}}%
    \gplfronttext
  \end{picture}%
\endgroup

\caption{The different determinations of the string tension as a
function of the number of smearing steps with $c=0.1$ for $L=23$ and $b=0.36$.
Errors reflect quality of the fit to Eq.~\ref{kappa}. }
%\end{figure}
}

Finally, we collect our results in Table~1. We write the number of
configurations, the value of $b$, $k$, $\bar{k}$ and $L$, and the result for the
string tension and plaquette expectation value. Statistical errors
were estimated by a jack-knife method, and are of order 5\%. A larger
fraction of the error is systematic. The results depends on the detail
of the fitting procedure (weight of the largest and smallest loops)
and in the number of smearing steps. For the $L=23$ case we have a
wider range of loop sizes with reasonably small errors and the
systematic errors are not bigger than 10\%. For the $L=17$ case the
statistics is higher, but this is well compensated by the limited
range of useful loop sizes. We also observed that at $\beta=0.36$ the
data presents large autocorrelation times. 
As emphasised  before, our goal here  is 
not to provide string tension determinations competitive with
others in the market, but rather to illustrate that very reasonable
results follow from the model. We draw the attention to how close our 
$L=23$ results are from the values   presented in
Ref.~\cite{LTW}  obtained from large $N$ extrapolation from 
standard lattice gauge theory determinations at  small values of $N$.
The $L=17$ results tend to be larger, but we find it premature, due to
the problems mentioned previously, to draw any conclusions from this 
fact. 

%\TABLE{
\begin{table}
\label{table1}
\begin{center}
\begin{tabular}{|| l | l | l | l| l | l | l||} \hline \hline
$N_{\rm conf}$ & $L$ & $b$  & $k$ & $\bar{k}$ & $\sigma$ & $\langle P_{\mu \nu} \rangle$ \\ \hline \hline
1025 & 17 & 0.36 & 5  & 7 & 0.055 (3)(6)  & 0.5581(1)\\ \hline
225 & 23 & 0.36 & 7  & 10 & 0.047 (3)(5)  & 0.5582(1)\\ \hline
225 & 23 & 0.37 & 7 &  10 & 0.035 (2)(4)  & 0.5789(1) \\ \hline  \hline
\end{tabular}
\end{center}
\caption{The values of the parameters of our simulations
together with the resulting value of the lattice string tension
$\sigma$ and the average value of the plaquette. 
The first error of the string tension is statistical and the second
systematic.}
\end{table}
%}

\section{Conclusions}
In this paper we have analysed the Twisted Eguchi Kawai (TEK) model in the weak
coupling phase for large group sizes $N=L^2$, and different values of
the  flux parameter $k$. Our result  confirms that fluxons, and in
particular torons(commuting fluxons),  seem to be the main mechanism driving
symmetry breaking for small $k$ values. This suggests that taking 
$k$ as a sufficiently large integer, and coprime with $L$, could solve
the instability of the model without sacrificing the perturbative
proof of reduction. Our numerical results confirm this expectation and 
show that for  $L/k<9$ no symmetry breaking is observed. Our
best choice is actually $L/k<4$ because in that case torons 
become unstable. It is also important to take into account that finite $L$
effects coming from the underlying non-commutative field theory could be 
made smaller by taking large values of $\bar{k}$ (defined in Eq.~\ref{defbark}),
even of order $L$. 

It is clear that there is still a lot to understand about the dynamics
of the TEK model at intermediate  couplings. A full semiclassical analysis 
including fluctuations around  fluxons is under way, and will be
included in a future publication~\cite{AGAMO4}. The possible evidence for a 
phase transition for  $k=(L-1)/2 $ (at $L=17,19,21,23$) suggest that other 
fluxons, or a different mechanism could be at work. At this stage it is 
difficult to determine if there is any relation  with the instabilities found in 
analytical~\cite{GHLL} and numerical works~\cite{BN} centred upon 
non-commutative field theories. It is obvious, however, that the 
twist-eater vacuum is the absolute minimum of the TEK model and cannot
present ``perturbative instabilities'',  but  these could be  hints of other sort
of problems observed numerically\cite{BN}.

Lacking, at this stage,  a proof that the symmetry breaking phase is
not metastable and that reduction will survive the
continuum limit, we decided to explore the present accessible ranges to
see whether physics results could be extracted from it. We showed that
one can access sizes which are much beyond $1/T_c$ and where
confinement behaviour is patent. The values obtained for the string
at our largest sizes $L=23$ scale in the right fashion.
Furthermore, they are surprisingly close to those obtained by
extrapolation from small groups~\cite{LTW}, given our limited
statistics. The $L$ dependence seems to be sizable, but we cannot
exclude that it is entirely due to our systematic errors. 
A more computer intensive study should be performed to pin
down differences, as well as $L$ and $k$ dependences. Anyhow, in our
opinion, we have shown evidence that the TEK model will continue to be
a powerful  tool to explore the dynamics of large $N$ Yang-Mills theory.

\section*{Acknowledgments}
The numerical simulations were made on Hitachi SR11000 computer at
Hiroshima University.

A. G-A want to express his gratitude to the Theoretical Physics Group
at Hiroshima University for the wonderful hospitality displayed to him
during the late stages of this work. In addition, A.G-A acknowledges
financial support from Comunidad Aut\'onoma
de Madrid under the program  HEPHACOS, the Spanish Ministry of Research 
through grants FPA2009-08785,  FPA2009-09017 and
Consolider-Ingenio 2010 CPAN (CSD2007-00042).

\bibliographystyle{JHEP}

\end{document}